\documentclass[fonts]{icst}

\usepackage{moreverb}

\usepackage[breaklinks,colorlinks,bookmarksopen,bookmarksnumbered,linkcolor=ICSTblue,citecolor=blue,urlcolor=ICSTblue]{hyperref}
\usepackage{breakurl}
\usepackage{doi}

\usepackage{graphicx}
\usepackage{epsf}
\usepackage{epsfig}
\usepackage{epstopdf}

\newcommand\BibTeX{{\rmfamily B\kern-.05em \textsc{i\kern-.025em b}\kern-.08em
T\kern-.1667em\lower.7ex\hbox{E}\kern-.125emX}}

\newtheorem{definition}{\bf Definition}

\def\volumeyear{2014}

\journalname{EAI Endorsed Transactions on Wireless Spectrum}
\articletype{Research Article/ICT.ORG}
 \def\volumeyear{2014}
 \def\volumenumber{01}
  \def\issuemonth{05-07}
  \def\issuenumber{2}
  \def\articleid{e2}
  \def\copyrightholder{Nima Namvar\textit{ et al.}, licensed to ICST}
  \copyrightnote{This is an open access article distributed under the terms of the Creative Commons Attribution license (\url{http://creativecommons.org/licenses/by/3.0/}), which permits unlimited use, distribution and reproduction in any medium so long as the original work is properly cited.}
  \def\articledoi{10.4108/ws.1.2.e2}
\received{18th Nov 2013}
  \accepted{13th Jun 2014}
  \published{14th Jul 2014}

\begin{document}

\runningheads{Nima Namvar \emph{et al.}}{Cell Selection in Wireless Two-Tier Networks: A Context-Aware Matching Game}

\title{Cell Selection in Wireless Two-Tier Networks: A Context-Aware Matching Game}

\author{Nima Namvar\affil{1}, Walid Saad\affil{2}, Behrouz Maham\affil{1}}

\address{\affilnum{1}Electrical and Computer Engineering Department,University of Tehran,Tehran, Iran.\\
\affilnum{2}Wireless@VT, Bradley Department of Electrical and Computer Engineering, Virginia Tech, Blacksburg, VA.}

\abstract{The deployment of small cell networks is seen as a major feature of the next generation of wireless networks. In this paper, a novel approach for cell association in small cell networks is proposed. The proposed approach exploits new types of  information extracted from the users' devices and environment to improve the way in which users are assigned to their serving base stations. Examples of such \emph{context} information include the devices' screen size and the users' trajectory. The problem is formulated as a matching game with externalities and a new, distributed algorithm is proposed to solve this game. The proposed algorithm is shown to reach a stable matching whose properties are studied. Simulation results show that the proposed context-aware matching approach yields significant performance gains, in terms of the average utility per user, when compared with a classical max-SINR approach.}




\maketitle

\section{Introduction}
Owing to the introduction of smartphones, tablets, and bandwidth-intensive wireless applications, the demand for the scarce radio spectrum has significantly increased in the past decade [1]. The concept of small cell networks (SCNs) is seen as a cost-effective and promising approach to cope with such an increasing demand. Indeed, the dense deployment of small cells, powered by low power, low cost base stations (BSs),  is seen as a promising technique to improve the coverage and capacity of wireless cellular systems [2-4]. However, due to the presence of different categories of cells with diverse power, capacity, and range, the introduction of such heterogeneous SCNs leads to many technical challenges such as resource allocation, network modeling, interference mitigation, and network economics [5].

One important challenge in SCNs is that of cell association and handover [6]. Indeed, developing approaches to assign mobile users to their preferred small cell while also handling prospective handovers is necessary to achieve efficient SCN operation. Due to the diversity of coverage-range of the cells in SCNs, applying traditional approaches for user-cell association (UCA) in an SCN can lead to undesirable network performance and possibly increased handover failures [7].

In [7], a user association algorithm based on traffic transfer is introduced which aims at pushing the users onto the more lightly loaded cells in order to improve load balancing in small cell networks. This is achieved by proposing a novel sub-optimal solution for optimizing the long-term rate that each user experiences. The authors in [8] propose a novel UCA strategy by joint optimization of channel selection and power control for the purpose of minimizing the delay. The authors use an approach that is related to the sum of per-user SINR. The work in [9] proposes a flexible UCA method which aims at reducing the outage probability of the network. This is done by analyzing the received SINR form each tier, when the tiers are distributed randomly according to Poisson process. A new approach for UCA in the downlink of small cell networks is introduced in [10] for increasing the minimum average users' throughput which is based on an iterative algorithm that exploits the feedback information of the users. The authors in [11] and [12] proposed a load-aware cell association strategy which, by adjusting the transmit power, dynamically modifies  the coverage area of the cells depending on their current load. This approach aims at balancing the load over neighboring macrocells. However, in small cell networks, one must balance the load over the various network tiers. A simple approach for user-cell association in small cell networks is proposed in [13]. In this approach, the authors use biasing factors for the transmit power of different tiers and attempt to distribute the traffic among the cells more fairly. Strategies based on channel borrowing from lightly-loaded cells are studied in [14-16]. In these works, some resources of lightly-loaded cells will temporarily be used for servicing the users in a neighboring cell. However such channel-borrowing strategies have been proposed for cell association in macrocell-only networks and are not effective in small cell networks. Other related works can be found in [17-20].

Most of this existing literature assumes that the network makes resource allocation and cell association decisions based solely on physical layer parameters. Indeed, the current state-of-the-art often ignores the fact that the users can have different mobility patterns and diverse quality-of-service (QoS) demands. However, an effective and optimum UCA approach must be able to distinguish the individual properties of the users and, thus, be able to prioritize them based on their traffic type (i.e. urgent real-time traffic and delay tolerant traffic), QoS demands, and trajectory. For instance, a fast-moving user that is using a video application should be treated differently from a semi-static user who is downloading a file. Here, the QoS of the first user could be dramatically impeded by the slightest of delays, while the latter is relatively delay tolerant. We refer to such additional information about the users or the network as \emph{context information}.

Thus, our main goal is to introduce a self-organizing approach for cell association in small cell networks, using which users and the network's cells can interact to decide on their preferred UCA in a way to optimize the overall network QoS. In particular, we propose a load-ware, application-aware approach for UCA which accounts for a plethora of context information including user mobility. Indeed, by exploiting context information from different network layers, we can develop a more efficient cell-association strategy which can lead to an improved network performance.

\begin{figure}
  \begin{center}
    \includegraphics[width=9cm]{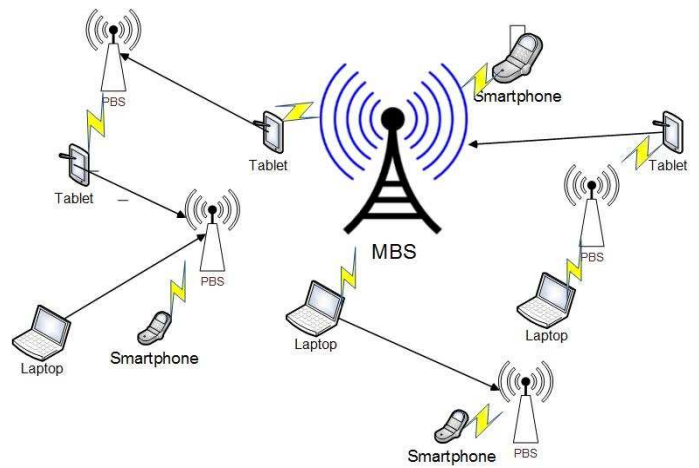}
   \caption{Users' mobility scenario in consideration}\vspace{-0.6cm}
  \end{center}
\end{figure}

The main contribution of this paper is to introduce a novel context-aware UCA approach which employs useful information from different features of the network in order to optimize the network-wide QoS. In our proposed model, we explore a combination of several context information which, to best of our knowledge, have not been used by any other work for user association in small cells: trajectory and speed of the users, cells' load, quality of service requirements of the users, and the hardware specification of the user equipments. We show that by utilizing the mentioned combination of context information, the network can better decide on which user should be assigned to which cell. We model the UCA problem as a many-to-one matching game with externalies. To do so, we introduce novel and well-defined utility functions to capture the preferences of the users and cells. To solve the proposed matching game, we propose a novel iterative algorithm that converges to a stable matching between the set of users and the set of the network's cells. Simulation results show that the proposed matching-based approach yields considerable QoS improvement relative to classical, context-unaware UCA approaches. The results also show that the proposed algorithm converges in  a reasonable number of iterations.

The rest of this paper is organized as follows: The system model is presented in Section 2. In Section 3, we formulate the user assignment problem in the framework of matching game with externalities and propose a novel algorithm to solve it. The performance of the proposed algorithm is assessed via simulations in the Section 4, and, finally, the conclusions are drawn in Section 5.

\section{System Model}\label{sec:sysmodel}\vspace{-0.1cm}
Consider the downlink of a two-tier wireless small cell network consisting of macrocells and picocells. Let $\mathcal{M}$, $\mathcal{P}$, and $\mathcal{N}$ denote the set of $\textit{M}$ macrocells, the set of $\textit{P}$ picocells, and the set of $\textit{N}$ users, respectively. Each small cell can serve a \emph{quota} of up  to $q$ users simultaneously. We assume a wireless channel having slow multipath fading. Users are moving at low speeds and request service from the different small cells that they meet during their travel in the network. Figure 1 shows a typical small cell network in which the users are mobile. As shown in Figure 1, the communication sessions should be handed over between the neighboring cells.

Each user in the network has its own performance indicators such as the urgency of data, and the QoS demand which depends on the hardware specification of a user's device and the application type. Thus, as a first step toward developing the proposed model, we will explicitly discuss all the user context information that will be accounted for.  

\underline{\textbf{Screen Size}}: The screen size of the user equipment will affect the QoS perception of the user, especially for video-oriented applications. Indeed, user equipments with large screens have more sensitive QoS perception to a video's resolution than the smaller user equipments. We capture the impact of the screen size of each user $i \in \mathcal{N}$ using a parameter $L_i$ that reflects the diameter length of each user's device. Devices with bigger screen size, are capable of showing the pictures with higher resolution which requires greater amount of network resources. Therefore, to satisfy the QoS demand of the devices with higher $L_i$, such as laptops or tablets, the network should allocate more resources to them relative to the smaller equipments such as smartphones.

\underline{\textbf{Data Urgency}}: The resource requirements of the users naturally depend on their traffic patterns and application requirements. For example, the QoS of a live video streaming vitally depends on the delivery time since a small amount of delay could decrease the QoS dramatically. In contrast, the download of an Internet file may not be too susceptible to delay. By prioritizing the users based on their QoS needs, we are able to improve the average QoS for the users while also distributing the traffic among the cells  more reasonably.

The QoS that each user experiences depends on the urgency of the user's data. Hence, we consider the QoS to be a function of delivery-time $t$. Naturally, for highly urgent data, the QoS will decrease more drastically as time elapses. Some suggestions to quantitatively model such behavior are presented in [21]. Consequently, for any user $i \in \mathcal{N}$, the QoS that reflects the data urgency can be given by:

\begin{equation}
Q_i(t) = \frac{1}{1+e^{t-\tau_i}},
\end{equation}
where $\tau_i$ is a parameter that reflects the urgency of the data. A smaller $\tau_i$ implies a more urgent data. This function shows that, within an interval of $2\tau_i$, the QoS drops to approximately $e^{-\tau_i}$ times of its initial value. This implies that only delivering the data before $\tau_i$ could be acceptable, and after that, the QoS becomes relatively small.


\textbf{\underline{Handover Process}}: Due to the mobility of the users, the active communication sessions must be handed over between the cells. Figure 1 shows the handover scenario in consideration. A handover (HO) process cannot occur immediately when a user enters to the boundary of the cell as it requires some initial preparation time. Prior to that, no data could be handed over between two neighboring cells. To guarantee the connection of the users to the cells, the network must avoid risky HOs that could potentially incur a signal loss or erroneous communication. A handover failure occurs when the received signal to noise and interference ratio (SINR) drops under a certain threshold [19]. Therefore, one can use received SINR to determine the handover-failure circles. In particular, we will use the typical value of -$6$ dB as the threshold of the received SINR for the handover-failure circle [20]. Here, we study the probability of handover failure (HF) considering the users' speed and trajectory. It is assumed that all cells are equipped with omnidirectional antennas. We assume a circular coverage area for tractability. We note that the matching approach presented in Section 3 can easily accommodate other forms of coverage areas and mobility models.

In a two-tier network, one must consider two handover types: 1) from macrocell to picocell and, 2) from picocell to picocell. Assume that a user that has previously been served by a macrocell enters a picocell submits a request for handover. When user $i\in \mathcal{N}$ enters a picocell $j\in \mathcal{P}$, the total possible time of interaction between the user and the picocell, $t_{T}^{ij}$, could be  computed as:
\begin{equation}\label{Intraction time}
  t_{T}^{ij}=\frac{2R_j\textrm{cos}(\theta_i)}{V_i},
\end{equation}
where $R_j$ represents the radius of the coverage area, and $\theta_i$ is the angle of the user's direction with respect to the imaginary line connecting it to the center of the cell as shown in Figure 2. $V_i$ is the user's average speed. Indeed, the numerator of (2) represents the length of the chord of the coverage circle that the user takes when it passes through the coverage area of the cell. Hereinafter, we assume that $V_i$ is small enough that channel conditions remains constant during the handover and that the users have \emph{low} to \emph{medium} mobility. A successful HO process necessitates a certain preparation time of duration $T_p$ before it could be initiated. Thus, based on the values of $t_T^{ij}$ and $T_p$, we distinguish two different scenarios: 1) If $t_T^{ij} > T_p$, the user is considered as a \emph{candidate} to be served; 2) If $t_T^{ij} < T_p$, the user is called a \emph{temporary guest} and no HO would be initiated.

\begin{figure}
  \begin{center}
    \includegraphics[width=9cm]{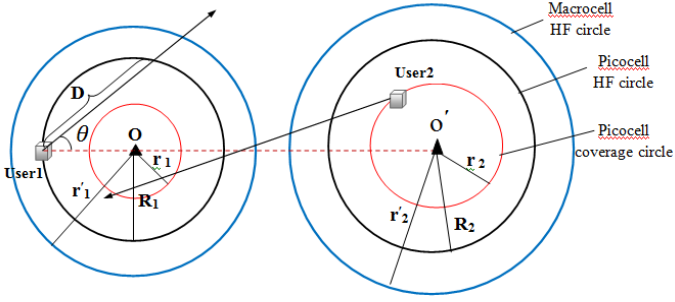}
   \caption{The handover failure and coverage regions}\vspace{-0.6cm}
  \end{center}
\end{figure}

The users enter the picocell at an arbitrary direction. Therefore, $\theta$ is a random variable which is distributed uniformly in $(-\frac{\pi}{2}, \frac{\pi}{2})$. Assume $D$ to be the length of the chord that the user takes. The cumulative distribution function (CDF) of D, $\textrm{Pr}(D<d)$, is equal to $2\textrm{Pr}(\theta>\textrm{cos}^{-1}\left(\frac{d}{2R}\right))$. Therefore, given that $\theta$ has uniform distribution, the probability density function (pdf) of $D$, $f_D(d)$, can be given by:

\begin{equation}
 f_D(d)=\frac{1}{\pi R\sqrt{1-\frac{d^2}{4R^2}}}.
\end{equation}

A handover process fails when the user's path intersects with the handover failure (HF) circle. When the path is the tangent of the HF circle (with the radius $r$), $D$ is equal to $2\sqrt{R^2-r^2}$. Therefore, when $D\geq2\sqrt{R^2-r^2}$, the user's path intersects with the HF circle and the handover fails. Using (3), the probability of HF when a user enters from macrocell to picocell (M2P) can be derived as follows:
\begin{equation}
  \textrm{Pr}_{HF}^{M2P} = \int_{2\sqrt{R^2-r^2}}^{2R}f_D(x)dx=\frac{2}{\pi}\textrm{cos}^{-1}\left(\sqrt{1-(\frac{r}{R})^2}\right).
\end{equation}
(4) shows that the probability of a handover failure is a function of $\frac{r}{R}$. Therefore, $\frac{r}{R}$ can be used as an indicator of the handover reliability. For example, assume that a handover could be initiated only if $Pr_{HF}(\frac{r}{R})\leq 0.05$; then the next cell must hold this condition: $\frac{r}{R}\leq 0.08$. If the cell does not satisfy this condition, then, no handover should be initiated. Indeed, he HO process becomes more reliable as $r$ becomes smaller relatively to $R$. The ratio of $r$ to $R$ varies from cell to cell and therefore, the different cells guarantee different levels of reliability during the handover process.

Now, assume that a user exits from picocell $j_1 \in \mathcal{P}$ and enters to another neighboring picocell $j_2 \in \mathcal{P}$ and sends a request for data handover. The handover process could be initiated once the user leaves $j_1$. However, it must be terminated before the user's distance from $j_1$ exceeds $r'_1>R_1$ and also before it enters the coverage of picocell $j_2$ to a distance of $r_2$. Let $O$ and $O'$ represent the centers of $j_1$ and $j_2$ respectively. Thus, $OO'$ represents the distance between the two picocell base stations. To ensure a reliable and successful handover, only those cells which satisfy the inequality $R_1+r_2 \leq OO'\leq r'_1+R_2$, must be considered for the handover.

The speed of the users can vary between two extremes $V_{min}$ and $V_{max}$. In practice, as the small cells often do not have all the information on the mobility distribution, then, it would be reasonable to assume that the users' speed varies uniformly between these two extents [24]. The probability of handover failure when a user enters from picocell to another picocell (P2P) can be computed by subtracting the probability of successful handover from (1). For a successful handover, two independent conditions must be satisfied. First, the user should move slowly enough so that the handover in the first cell could be triggered. The probability of this event is given by $\textrm{Pr}(V<\frac{r'_1-R_1}{t_{m_1}})$. Second, the path of the user should be in such a way that it does not intersect with the HF circle of the destination cell. Therefore, given that users' speed has a uniform distribution, the probability of handover failure is given by:

\begin{multline}\
  \textrm{Pr}_{HF}^{P2P}
  =1-\frac{\frac{r'_1-R_1}{T_{p_1}}-V_{min}}{V_{max}-V_{min}}\left(1-\frac{2}{\pi}\textrm{cos}^{-1}(\sqrt{1-(\frac{r_2}{R_2})^2})\right).
\end{multline}

Now, considering the defined context information, in the next section, we formulate the UCA problem as a context-aware many-to-one matching game.

\section{Cell Association as a Matching Game with Externalities}\vspace{-0.1cm}
Originally introduced by Gale and Shapley in their seminal work [25], matching games are seen as a powerful and efficient framework to model conflicting objectives between two sets of players. Players of each set have a ranking, or preference, over the players in the opposite set. These preferences capture the objectives of players and the purpose of a matching game is to match the players of these two sets according to their preferences [26].

Among different types of matching games, the many-to-one matching scenario is especially suitable for the studied cell association problem because in this game, several players of one set can be matched with a single player of the other set. As an analogy to the many-to-one matching game, in the cell association problem several users can be assigned to a single cell. Here, using the context information introduced in the previous section, we can define proper utility functions to capture the preferences of users and small cells. Once this is done, the many-to-one matching model could be employed to assign the users to the cells based on each player's individual preferences and goals. In other words, using many-to-one matching games, we aim at maximizing the utility functions of users and small cells and thereby, optimizing the network-wide performance.

In the classical matching game introduced in [25-27], it is assumed that the preferences of the players are independent. However, this assumption does not hold in our model since the QoS metrics of the players are interdependent. In other words, as we can see from (6) and (7), the prospective utilities of the cells and users must depend on the current matching which itself depends on the preferences of the players. In such situations in which externalities affect the preferences of the players, the many-to-one matching game model with externalities is a promising approach to study the problem [28], [29]. However, there is no general solution for matching games with externalities as the general approach of Gale and Shapley cannot be generalized to this case. Therefore, introducing a novel approach which is tailored to specific nature of the proposed game is required. Indeed, the unique properties of our problem requires the introduction of a novel solution to the matching game which is tailored to the specific nature of the UCA problem.


Formally, the outcome of the UCA problem is a \textit{matching} between two sets $\mathcal{N}$ and $\mathcal{P}$ which is defined as follows:

\begin{definition}
A \emph{matching} $\mathcal{\mu}$ is a function from $\mathcal{N}\cup \mathcal{P}$ to $2^{\mathcal{N}\cup \mathcal{P}}$ such that $\forall n\in \mathcal{N}$ and $\forall p\in \mathcal{P}$: (i) $\mu(n)\in \mathcal{P}\cup \emptyset$ and $|\mu(n)|\leq1$, (ii) $\mu(p)\in 2^{\mathcal{N}}$ and $|\mu(p)|\leq q_p$, and (iii) $\mu(n)=p$ if and only if $n$ is in $\mu(p)$.
\end{definition}

The users who are not assigned to any member of $\mathcal{P}$, will be assigned to the nearest macrocell. Members of $\mathcal{N}$ and $\mathcal{P}$  must have strict, reflexive and transitive preferences over the agents in the opposite set. 
In the next subsections, exploiting the context information we introduce some properly-defined utility functions to effectively capture the preferences of each set.


\subsection{Users' Preferences}\vspace{0.1cm}
Each user seeks to maximize its QoS requirements. Indeed, the users prefer those cells that are able to provide a reasonable delay while also meeting the QoS requirements as dictated by the application type and the screen size of each user's device. Users require a target rate $\hat{C}$ that reflects the type of applications which fits their screen size. Therefore, for each user $i\in \mathcal{N}$ with screen size $L_i$, we assign a target rate $\hat{C}_i(L_i)$ which quantifies the QoS requirement of the user. Moreover, the users seek to optimize their transmission rate which depends on the received power and the interference caused by neighboring small cells. Hence, those cells that are less congested and have higher transmission rate are prioritized by the users. In fact, the available amount of resources in a cell depends on the number of its current users, in such a way that the less congested the cell is, the more resources could likely be available. For each user $i$ serviced by a small cell $j$, the utility function can be given by:

\begin{equation}
 U_i^{user}(\mu,j,L_i )=
\begin{cases}
(\frac{C_i-\hat{C_i}(L_i)}{K_i})^{\alpha_i}-\gamma_i(q_j-m_j)& \\ \text{if } \hat{C_i}(L_i)\leq C_i, \\ \text{ }
\\-\lambda_i\left(\frac{\hat{C_i}(L_i)-C_i}{K_i}\right)^{\beta_i}-\gamma_i(q_j-m_j) & \\ \text{if } \hat{C_i}(L_i)>C_i,\\
\end{cases}
\end{equation}
where $q_j$ is the quota of the small cell $j$, and $m_j$ is the total number of users being served by it. $L_i$ is the screen size of user $i$ and $\hat{C}_i$ is the its target rate. $C_i$ represents the received rate of the user $i$ which is equal to $W\textrm{log}_2(1+\frac{P_jc_{ij}}{\sum_{k \neq j} P_kc_{ik}+\sigma^2})$, where $P_j$ is the power of small cell base station (SCBS) $j$, $c_{ij}$ is the channel coefficient between user $i$ and SCBS $j$, $\sigma^2$ is the power of additive noise, and $W$ is the bandwidth. $\gamma_i$ is the cost per unit traffic and $\alpha_i$, $\beta_i$, $\lambda_i$ and $K_i$ are the coefficients that shape the utility function.





Figure 3 shows an example of the utility of a user for $\gamma=0$. This illustrative example will show how each user, having different screen size, can perceive the rate gains. As we can see, for large-screen devices, such as laptops, the utility of the users is very sensitive to the received rate since a large screen allows users to better discern the quality of the application being used (e.g. video or multimedia). In contrast, the utility of the users with small screen size is not too susceptible to the received rate. Therefore, users on smartphones will overweight low rates (with respect to the reference $\hat{C}$), since the quality might be perceived as good, even though in reality it is below par. Moreover, because they are not capable of showing the pictures with extremely high resolution, receiving rates that are much higher than the target rate cannot change the utility of users with small screens significantly.

\begin{figure}
  \begin{center}
    \includegraphics[width=9.5cm]{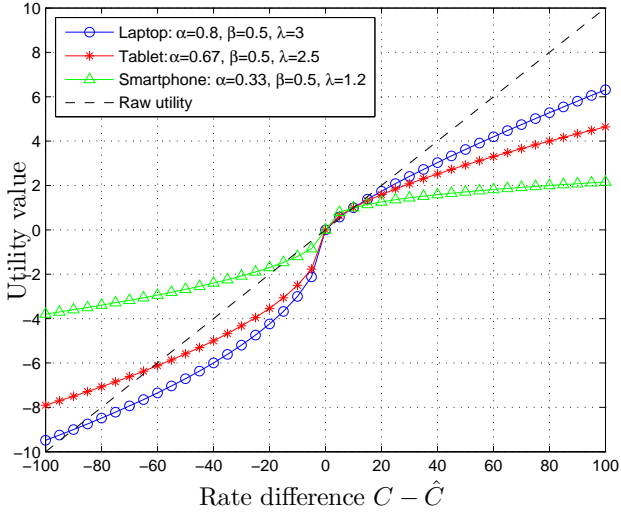}
    \vspace{-0.5cm}
   \caption{Utility of the users with different screen size}\vspace{-0.6cm}
  \end{center}
\end{figure}

The value of $m_j$ depends on the current matching, because it is the current matching that determines how many users are assigned to a specific small cell. As a result, the utility of each user is a function of current matching $\mu$, as shown in (6). The first term in (6) captures the user's natural objective to maximize its transmission rate and the second term accounts for the fact that the users seek to find lightly loaded small cells to achieve more resources.

In fact, this utility function encourages the user to select lightly loaded cells and consequently, helps to offload the heavily-loaded cells by pushing the users to more lightly-loaded cells. Using (6), the users can rank the SCBSs in their vicinity based on the defined utility.



\subsection{Small Cells' Preferences}
The main goal of each small cell is to increase the network-wide capacity by offloading traffic from the macrocells while providing satisfactory QoS for the users. To decrease the number of total handovers, the small cells prefer the users which stay longer in the cell. The possible interaction time between the user and the cell depends on the speed and direction of the user. Clearly, users with lower mobility and a trajectory close to the cell's diameter would stay longer in the cell. On the other hand, to increase the network-wide QoS, the small cells must prioritize users having more urgent requests compared to those with less urgent ones.


By prioritizing the users coming from congested cells, the small cells could offload the heavily-loaded cells. To encourage the cells to prioritize the users coming from congested cells, we assume that each user is carrying a potential utility as a function of the pervious cell $j'$ load, $f(\frac{m_{j'}}{q_{j'}})$. This utility depends on the current matching which determines the number of users in neighboring cells. We define the following utility that each SCBS $j \in \mathcal{P}$ obtains by serving an acceptable UE $i\in \mathcal{N}$:
\begin{align}
U_{j}^{SCBS}\left(\mu,i,m_{j'},q_{j'}\right)= \nonumber \\
 \frac{\textrm{cos}(\theta_i)}{V_i} \left[1+\textrm{log}\left(\frac{\textrm{max}(1,m_{j'})}{q_{j'}}\right)\right] \frac{1}{\tau_i}.
\end{align}

The first term in (7) allows to prioritize the users that stay longer in the cell. The second term accounts for the offloading concept, and the third term is the utility achieved by the SCBS $j$ when serving a specific application. This utility function is well matched with the fact that a given small cell gains more utility by giving service to the users that are moving slower, having more urgent data, and coming from more congested cells. Thus, by doing so, the network could provide higher QoS and distribute the load more effectively.


From (6) and (7), we can see that the utilities depend on the current matching $\mu$ and consequently, the preferences of the players are interdependent. Under this condition, the preferences of players are not solely based on individuals, but some \emph{externalities} affect the preferences and matching as well.


\begin{definition}
The preference relation $\succ_i$ of the user $i\in \mathcal{N}$ over the set of matchings $\Psi(\mathcal{N},\mathcal{P})$ is a function that compare two matchings $\mu, \mu' \in \Psi$ such that:
\begin{equation}
  \mu \succ_i \mu' \Leftrightarrow U^{user}_i(\mu,j,L_i)>U^{user}_i (\mu',j,L_i).
\end{equation}
\end{definition}

The preference relation for an SCBS $j$, $\succ_j$, is defined similarly. Users and SCBSs rank the members of the opposite set based on the defined preference relations. Our purpose is to match the users to the small cells so that the preferences of both side are satisfied as much as possible; thereby the network-wide efficiency would be optimized.

To solve a matching game, one suitable concept is that of a stable matching. In a matching game with externalities, stability has different definitions based on the application. Here, we consider the following notion of stability: 

\begin{definition}
A matching $\mu$  is blocked by the user-SCBS pair (i,j) if $\mu(i)\neq j$ and if $j\succ_i \mu(i)$ and $i\succ_j i'$ for some $i'\in \mu(j)$. A many-to-one matching is \emph{stable} if it is not blocked by any user-SCBS pair.
\end{definition}

In the next section, we propose an efficient algorithm for solving the game that can find a stable matching between users and small cells.

\begin{table}[t] 
  \scriptsize
  \centering
  \caption{Proposed Algorithm For The Matching Game}
    \begin{tabular}{p{8.5cm}}
      \hline
      \vspace{0.1cm}
\textbf{Input:} context-aware utilities and the preferences of each set \\
\textbf{Output:} Stable matching between the users and SCBSs \\
\\
\textbf{Initializing}: All the UEs are assigned to the nearest macro-BS \\
\\
\textbf{Stage I}: \textbf{Preference Lists Composition}
\begin{itemize}
  \item UEs and SCBSs exchange their context information
  \item UEs(SCBS) sort the set of acceptable candidate SCBSs(UEs) based on their preference functions
\end{itemize} \\
\textbf{Stage II}: \textbf{Matching Evaluation} \\
\hspace*{1em}\textbf{while:} $\mu^{(n+1)}\neq \mu^{(n)}$
\begin{itemize}
  \item Update the utilities based on the current matching $\mu$
  \item Construct the preference lists using preference relations $\succ_i$ and $\succ_j$ for $\forall i\in \mathcal{N} \text{ and } \forall j\in \mathcal{P}$
  \item Each user $i$ applies to its most preferred SCBS
  \item Each SCBS $j$ accept the most preferred applicants up to its quota $q_j$ and create a waiting list while rejecting the others
\end{itemize}
\hspace*{2em} \textbf{Repeat}\\
\hspace*{4em} $\bullet$ Each rejected user applies to its next preferred SCBS \\
\hspace*{4em} $\bullet$ Each SCBS update its waiting list considering the new applicants \\
\hspace*{5em} and the pervious awaiting applicants up to its quota\\
\hspace*{2em} \textbf{Until:} all the users assigned to a waiting list \\
\hspace*{1em}\textbf{end} \\
   \hline
    \end{tabular}\label{tab:algo}

\end{table}

\subsection{Proposed Algorithm}
The deferred acceptance algorithm, introduced in [26], is a well-known approach to solving the standard matching games. However, in our game, the preferences of the players as shown in (7) and (9), depend on externalities through the entire matching, unlike classical matching problems. Therefore, the classical approaches such as deferred acceptance cannot be used here because of the presence of externalities [28],[29]. To solve the formulated game, we propose a novel algorithm shown in Table I. Assume that all the users are initially associated to the nearest macro base station (MBS). Each user sends its profile information ($V$, $\alpha$, $\tau$) to the neighboring SCBSs. Each SCBS, on the other side, only keeps the users satisfying (8) and ranks them based on their utilities (9). After ranking the acceptable UEs, the SCBS sends to the currently waiting users its own context information including its rate over load defined in (6) and its corresponding coverage and HF circle radii $R$ and $r$.

Each user makes a ranking list of the available SCBSs and applies to the most preferred one. The SCBSs rank the applying users and keep the most preferred ones up to their quota and reject the others. The users who have been rejected in the former phase, would apply to their next preferred SCBS and the SCBSs modify their waiting list accordingly. This procedure continues until all the users are assigned to a waiting list.

\begin{table}
\caption{Typical values of data rate for different devices}
\centering

\begin{tabular}{c|c|c}

\hline
Device type & Average screen size & Typical Data rate\\
\hline
Laptop     & 17''   & 1000 kbps\\
Tablet     & 10''  & 600 kbps\\
Smartphone & 4.5''    & 400 kbps\\
\hline

\end{tabular}
\end{table}

 However, since the preferences depend on the current matching $\mu$, an iterative approach should be employed. In each step, the utilities would be updated based on the current matching. Once the utilities are updated, the preference lists would be updated accordingly as well. Therefore, in each iteration, a new temporal matching arises and based on this matching, the interdependent utilities are updated as well. The algorithm initiates the next iteration based on the modified preferences. The iterations will continue until two subsequent temporal matchings are the same and algorithm converges.

 The proposed algorithm will lead to a stable matching when it converges, since by contradiction, the ``deferred acceptance'' in Stage II would not converge if the matching is not stable. Although a formal analytical proof of convergence for the proposed algorithm is difficult to derive, we make several observations that can help in establishing such a convergence. First, we note that in each iteration the ``deferred acceptance'' method in Stage II yields a temporary matching between the users and cells for any initial preferences [25], [26]. Following each iteration, the preferences are updated according to (5) and (6) which are functions of three main variables: the topology and speed of users, the channel conditions, and the current matching.

Second, in view of the fact that users have low mobility and experience a wireless channel with slow fading, we can assume that the network's topology and channel conditions remain almost constant during an algorithm run. As a result, we can conclude that in each iteration the preferences are updated solely based on the current temporary matching. Therefore, since there is only a finite number of possible matchings between the users and their neighboring cells, the updating the preferences is not an endless process. In other words, there would be a limited number of iterations which beyond that, updating the preferences will either converge to a final, stable matching or cycle between a number of temporary matchings. However, here, we note two things: a) based on our thorough simulation results in Section 4, the case in which there is a cycling behavior only rarely occurs and b) under this case,  we assume that the players can detect a cycle and stop the algorithm.

\section{Simulation Results}
For our simulations, we consider a single MBS with radius 1 km and  overlaid by $P$ uniformly deployed picocells. The transmit power of each picocell is $30$ dBm and its bandwidth is $W=200$ kHz. The small cells' quota is supposed to be a typical value $q=$4 for all SCBSs [30]. The channels experience a Rayleigh fading with parameter $k=2$. Noise level is assumed to be $\sigma^2=-121$~dBm and the minimum acceptable SINR for the UEs is 9.56 dB [31]. 
There are $N$ users distributed uniformly in the network. The QoS parameter $\tau_i$ in (1) is chosen randomly from the interval [0.5,5] ms. The users have low mobility and can be assumed approximately static during the time required for a matching. The speed of users varies between $20 km/h$ and $40 km/h$. Utility parameters in (6) are chosen in line with Figure 3. $\gamma_i$ and $K_i$, are assumed to be $1$ and $10$ respectively, for all the users $i \in \mathcal{N}$. All the statistical results are averaged via a large number of runs over the random location of users and SCBSs, the channel fading coefficients, and other random parameters. The performance is compared with the max-SINR algorithm which is a well-known context-unaware approach used in wireless cellular networks for the UCA. In this approach, each user is associated to the SCBS providing the strongest SINR.

Figure 4 shows the average received rate per user for different number of SCBSs. As the number of SCBSs increases, the interference between the different cells increases. Therefore, the average rate that each user achieves will decrease. Figure 4 demonstrates that the proposed algorithm can lead to higher average rate per user in comparison to max-SINR approach reaching up to $66.7\%$ gain for a network size of $P=36$ SCBSs.

\begin{figure}\vspace{-1em}
  \begin{center}
    \includegraphics[width=9.5cm]{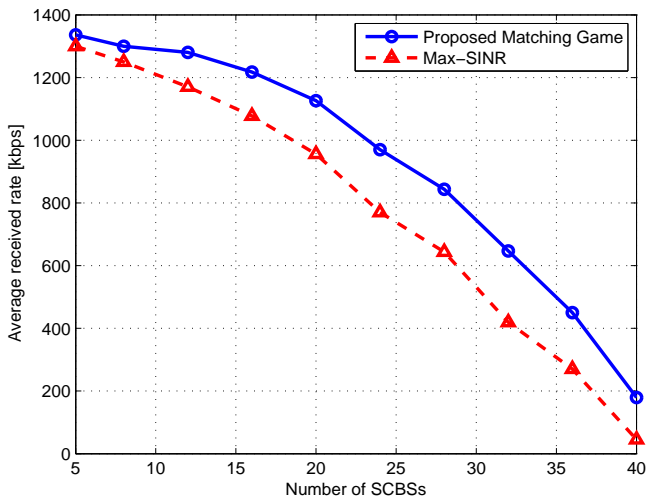}
    \vspace{-0.5cm}
     \caption{Average received rate per user for different number of SCBSs with $N=60$ users.}\vspace{-2em}
  \end{center}
\end{figure}\vspace{-0.1em}

Figure 5 shows the average utility per different types of devices, for different number of SCBSs when the number of users is $N = 60$. According to (6), each user has a specific target rate tailored to its screen size. Typical values used for the target rates for three different types of devices are shown in Table 2. Figure 5 shows that, for small-screen devices such as smartphones, the perceived utility of the user will not change dramatically if it receives a rate that is higher than its target rate. However, this utility for larger devices such as tablets and laptops is more sensitive to the received rate. From Figure 5, we can see that, when the number of SCBS is small and the average received rate is high, the utility of the laptops and tablets is greater than that of the smartphones because they are more sensitive to the received rate. However, as the number of the SCBSs increases and the network becomes more congested, the average received rate decreases and the utility of laptops and tablets decreases considerably, while the utility of the smartphones decreases very slowly. In Figure 5, we can see that, in general, for all types of devices, the proposed approach outperforms the conventional max-SINR approach.

Figure 6 shows the average utility per user for different number of SCBSs for $N = 60$ users. As the number of SCBSs increases, the average utility per user will decrease because the received rate will decrease due to the stronger interference. Although the cost for the traffic will also decrease (second term in (6)) when the number of SCBSs increases, but its effect is less than the effect of rate (first term in (6)). Figure 6 shows that the proposed algorithm outperforms the max-SINR algorithm for all network sizes. This performance advantage reaches up to $194\%$ gain over to max-SINR criterion for a network with $24$ SCBSs.

Figure 7 shows the average utility per user for different types of devices and, for different number of users when the number of SCBSs is $P = 15$. In Figure 7, we can see that, as the number of users increases, the average received rate per user will also increase. Therefore, the utility of the devices which is a function of the received rate will increase as well. However, when the average received rate is small, devices with smaller screens have more utility relative to the ones with large screens. This is due to the fact that the small devices are not so sensitive to the rate since they are incapable of handling higher resolutions. Similar to Figure 5, in Figure 7, we can see that devices with larger screen size are more susceptible to the received rate, i.e. the distance from the BS. In fact, as the rate increases, we can see that the devices with large screen size such as laptop, achieve more utility in comparison to the small devices, since they are so sensitive to the rate and an increase in the received rate can increase their QoS considerably. We can see from Figure 7 that the proposed algorithm has noticeable gain over the max-SINR approach and can reach up to $4\%$, $32\%$, and $87.5\%$ gain over the max-SINR criterion for the smartphones, tablets, and laptops respectively.

\begin{figure}\vspace{-1em}
  \begin{center}
    \includegraphics[width=9.5cm]{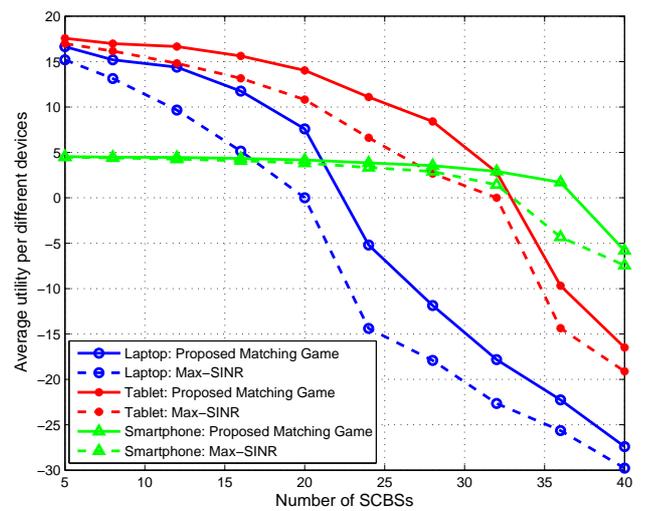}
    \vspace{-0.7cm}
     \caption{Average utility per different types of devices with $N=60$ users.}\vspace{-2em}
  \end{center}
\end{figure}\vspace{-0.1em}

\begin{figure}\vspace{-1em}
  \begin{center}
    \includegraphics[width=9.5cm]{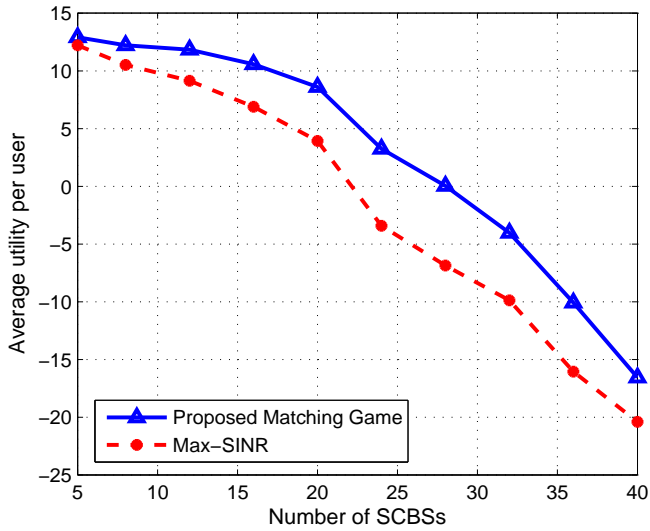}
    \vspace{-0.5cm}
     \caption{Average utility per user for different number of SCBSs with $N=60$ users.}\vspace{-2em}
  \end{center}
\end{figure}\vspace{-0.1em}

Figure 8 shows average utility per user for different number of users with $P = 15$ SCBSs. As the number of users increases, the average received rate will also increase which leads to an increase in the average user's utility. Figure 8 demonstrates that at all network sizes, the proposed approach has a performance advantage over max-SINR. The average gain of the proposed approach over the  max-SINR scheme is $39.4\%$.

Figure 9 shows the average utility per user for different percentage of the smartphones for a network size of $N=60$ users and $P=20$ SCBSs. As the percentage of the smartphones increases from  $50\%$ to $100\%$, the gain of the proposed approach relative to max-SINR scheme decreases from  $113\%$ to $9\%$. This is directly related to the features of the smartphones. In fact, devices with small screen size are not very sensitive to the received rate, therefore, the proposed context-aware UCA algorithm which aims at optimizing the received rate of the devices will not have considerable gains over the context-unaware max-SINR approach when the network encompasses devices with small screens only. Conversely, when the network has considerable percentage of laptops and tablets which are very sensitive to the received rate, then the proposed context-aware approach yields significant gain over the max-SINR because the proposed algorithm prioritize the devices based on their QoS demands and requirements.

In Figure 10, we show the average utility achieved by each SCBS as a function of the number of users for $P=15$ SCBSs. As the number of users $N$ increases, the network becomes more congested, and the probability that a new user who applies for an SCBS is coming from a congested BS increases. Therefore, it is more likely for the SCBSs to gain more utility by offloading the network. However, when the network is considerably congested, the new users that arrive to the network would be mostly assigned to the MBS, since many of SCBSs have already reached their maximum capacity. Figure 10 shows that, at all network sizes, the proposed algorithm achieves significant gains over the max-SINR approach that reach up to $72.8\%$ gain for a network size of $40$.

Figure 11 shows the average number of iterations per user required for the algorithm to converge to a stable matching for two different network sizes, as the number of users varies. In this figure, we can see that the number of algorithm iterations is an increasing function of the number of users and the number of SCBSs. Figure 10 shows that the average number of iterations varies from $1.09$ and $1.1$ at $N=3$ to $8.3$ and $9.7$ at $N=80$, for the cases of $15$~SCBSs and $20$ SCBSs, respectively. Clearly, Figure 11 demonstrates that the proposed algorithm converges within a reasonable number of iterations and scales well with the network size.

\begin{figure}\vspace{1em}
  \begin{center}
    \includegraphics[width=9.5cm]{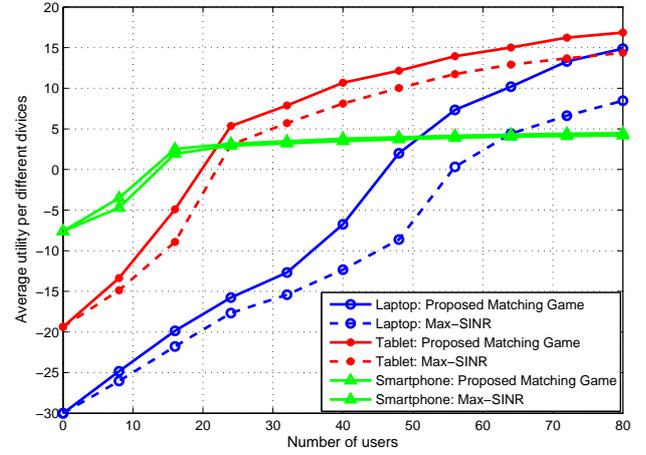}
    \vspace{-0.3cm}
     \caption{Average utility per different types of devices with $P=15$ SCBSs.}\vspace{-1.75em}
  \end{center}
\end{figure}\vspace{-0.1em}

\section{Conclusions}
In this paper, we have proposed a new context-aware user association algorithm for the downlink of wireless small cell networks. By introducing well-designed utility functions, our approach accounts for the trajectory and speed of the users as well as for their heterogeneous QoS requirements and their hardware specifications. We have modeled the problem as a many-to-one matching game with externalities, where the preferences of the players are interdependent and contingent on the current matching. To solve the game, we have proposed a novel algorithm that converges to a stable matching in a reasonable number of iterations. Simulation results have shown that the proposed approach yields considerable gains compared to max-SINR approach.

\section{Acknowledgements}
This work was supported by the U.S. National Science Foundation under Grant CNS-1253731.

\begin{figure}\vspace{-1em}
  \begin{center}
    \includegraphics[width=9.5cm]{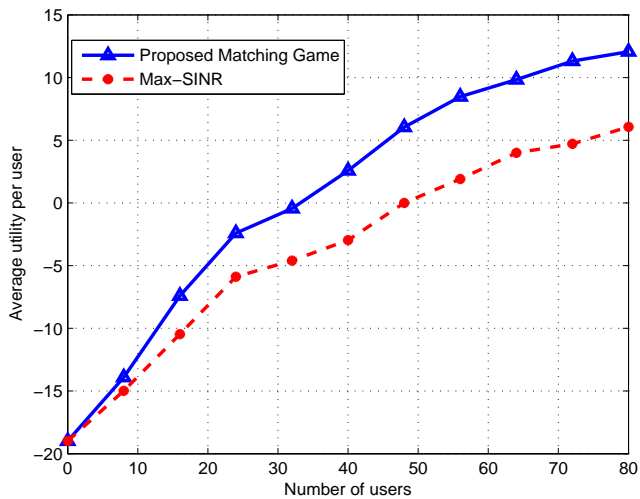}
    \vspace{-0.5cm}
     \caption{Average utility per user for different number of users with $P=15$ SCBSs.}
  \end{center}
\end{figure}\vspace{-0.1em}

\begin{figure}\vspace{-3em}
  \begin{center}
    \includegraphics[width=9.5cm]{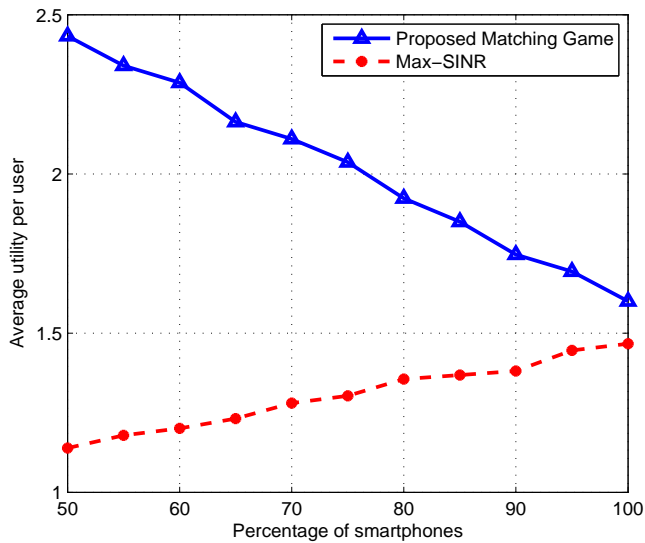}
    \vspace{-0.5cm}
     \caption{Average utility per user for different percentage of the smartphones with $N=60$ users and $P=20$ SCBSs.}
  \end{center}
\end{figure}\vspace{-0.1em}

\begin{figure}\vspace{-1em}
  \begin{center}
    \includegraphics[width=9.5cm]{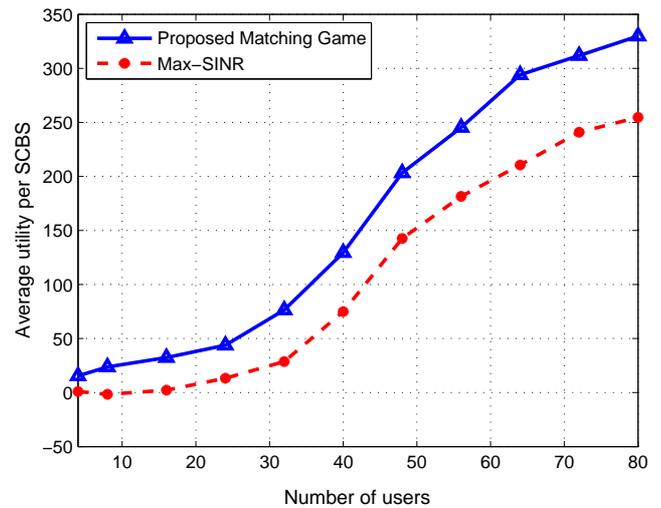}
    \vspace{-0.4cm}
     \caption{Average utility per SCBS for different number of users with $P=15$ SCBSs.}
  \end{center}
\end{figure}\vspace{-0.1em}

\begin{figure}\vspace{-1em}
  \begin{center}
    \includegraphics[width=9.5cm]{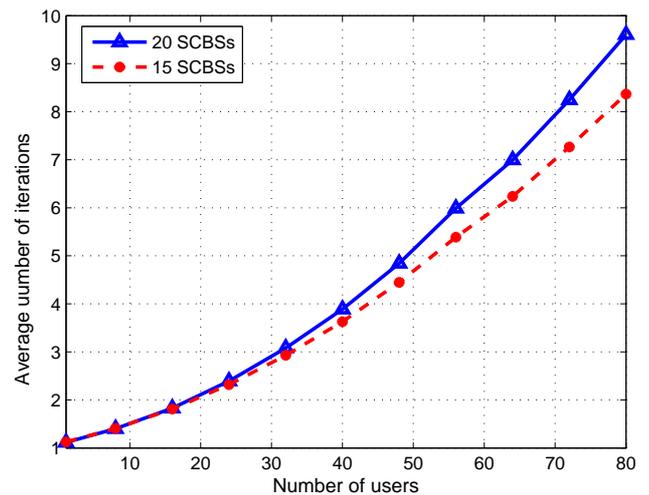}
    \vspace{-0.5cm}
     \caption{Average number of algorithm iterations for reaching a stable matching, for different number of users with $P=15$ and $P=20$ SCBSs.}
  \end{center}
\end{figure}
\end{document}